\documentclass[11pt, a4paper]{article}
\pdfoutput=1

\usepackage[affil-it]{authblk}
\usepackage[latin1]{inputenc}
\usepackage{amsmath}
\usepackage{amsfonts}
\usepackage{amssymb}
\usepackage{latexsym}
\usepackage{mathrsfs}
\usepackage{style}
\usepackage{graphicx}
\usepackage{bm}
\usepackage{amsthm}

\def\M{\mathcal{M}}

\title{Soft sub-leading divergences in Yang-Mills amplitudes}

\author{Eduardo Casali%
  \thanks{Email: \texttt{e.casali@damtp.cam.ac.uk}}}
\affil{Department of Applied Mathematics and Theoretical Physics,\\ Wilberforce Road, Cambridge CB3 0WA, UK}
\date{}
\begin{document}

\maketitle
\begin{abstract}
In this short note I show that the soft limit for colour-ordered tree-level Yang-Mills amplitudes contains a sub-leading divergent term analogous to terms found recently by Cachazo and Strominger for
tree-level gravity amplitudes. 

\end{abstract}

\section{Introduction}

Recently soft limits have been investigated from the point of view of asymptotic symmetries of flat space \cite{Strominger:2013lka,Strominger:2013jfa,He:2014laa}, in particular Weinberg's soft graviton theorem has been
derived from the action of the super-translations of the BMS group on the gravitational S-matrix. The BMS group can be enlarged by considering transformations that act singularly on the sphere at infinity \cite{Barnich:2011ct,Barnich:2011mi},
these super-rotations will act on the S-matrix and it has been conjectured that they might be connected to a new universal sub-leading divergence around the soft-limit for gravitational scattering amplitudes \cite{Cachazo:2014fwa}.
With that in mind I derive the analogous sub-leading divergence for colour-ordered Yang-Mills amplitudes and show that they are given by an angular momentum generator acting on the scattering amplitude, it is not unreasonable to
conjecture that this sub-leading term might be connected to symmetries of the asymptotic data of Yang-Mills on flat spacetime.

The usual soft limit is taken by introducing a small parameter $\epsilon$ which re-escales one of the gluons
\begin{equation}
 \lam_s\rightarrow\sqrt{\eps}\lam_s,\;\;\tilde{\lam}_s\rightarrow\sqrt{\eps}\tilde{\lam}_s
\end{equation}
but in order to see the sub-leading term a related limit must be used. The holomorphic version of the soft-limit \cite{ArkaniHamed:2008gz} is related to the usual one by a little group transformation
\begin{equation}
 \M(\{\sqrt{\eps}\lam,\sqrt{\eps}\tilde{\lam},+1\})=\eps\M(\{\eps\lam,\tilde{\lam},+1\}).
\end{equation}
Notice that this relation differs from the one from gravity by a power of $\eps$, so in Yang-Mills there's no sub-sub-leading divergence as in gravity, the claim is then
\begin{equation}
 \M_{n+1}(\{\eps\lam_s,\tilde{\lam}_s\},1,\dots,n)=\left(\frac{1}{\eps^2}S^{(0)}+\frac{1}{\eps}S^{(1)}\right)\M_n(1,\dots,n)+\O(\eps^0)
\end{equation}
Where $S^{(0)}$ is the usual soft limit for colour ordered YM amplitudes, and 
\begin{equation}
 S^{(1)}=\frac{E_\nu q_\mu J^{\mu\nu}_a}{q\cdot k_a}
\end{equation}
when the soft particle is adjacent to particle $a$. Notice that this term is gauge-invariant due to the antisymmetry of $J^{\mu\nu}$ so it has more in common with what was called $S^{(2)}$ in \cite{Cachazo:2014fwa}.
In spinor notation this can be written as
\begin{equation}
 \frac{E_\nu q_\mu J^{\mu\nu}_a}{q\cdot k_a}=\frac{\mu_\alpha\tilde\lam_{s\dot\alpha}}{\bra\mu s\ket}\frac{\lam_{s\beta}\tilde\lam_{s\dot\beta}}{\bra s a\ket\sbra s a\sket}(\varepsilon^{\al\beta}\tilde J^{\dot\al\dot\beta}+\varepsilon^{\dot\al\dot\beta}J^{\al\beta})
\end{equation}
where
\begin{equation}
 J^{\al\beta}=\lam^\al\frac{\pd}{\pd\lam_\beta}+\lam^\beta\frac{\pd}{\pd\lam_\al},\;\;\;\tilde J^{\dot\al\dot\beta}=\tilde\lam^{\dot\al}\frac{\pd}{\pd\tilde\lam_{\dot\beta}}+\tilde\lam^{\dot\beta}\frac{\pd}{\pd\tilde\lam_{\dot\al}}
\end{equation}

\section{Proof}

The proof follows closely the one given in \cite{Cachazo:2014fwa} and I adopt their notation. Consider an $n+1$ particle, colour-ordered amplitude in Yang-Mills at tree-level,
$\mathcal{M}_{n+1}(s,1,\dots,n-1,n)$, where s is the soft particle which for simplicity I'll take it to have helicity $h_s=+1$. The amplitude without the momentum conserving delta function will be called $M$,
deform this stripped amplitude using the BCFW shift
\begin{equation}
 \lam_s(z)=\lam_s+z\lam_n,\;\;\tilde{\lam}_n(z)=\tilde{\lam}_n-z\tilde\lam_s,
\end{equation}
then the amplitude factorizes as
\begin{equation}
 M_{n+1}=\sum M_L(s(z^*),1,\dots,j,I)\frac{1}{P^2_I}M_R(-I,j+1,\dots,n(z^*))
\end{equation}
where the sum is over the set of ordered particles and helicities of the internal particle.

In the soft limit the only interesting term in this sum is when $j=1$, where $M_L$ is a three particle amplitude. The other terms are finite under the soft limit, the proof is the same as the one
given on appendix A of \cite{Cachazo:2014fwa}. Therefore in the following I'll drop all other terms except
\begin{equation}
 M_L(s(z^*),1,I)\frac{1}{P^2_I}M_n(-I,2,\dots,n-1,n(z^*)).
\end{equation}

Finding the pole gives
\begin{equation}
 (k_s(z^*)+k_1)^2=0,\;\;z^*=-\frac{\bra 1 s\ket}{\sbra 1 n\sket}
\end{equation}
which fixes the internal spinor to be
\begin{equation}
 \lam_I=\lam_1,\;\;\;\tilde\lam_I=\frac{\bra ns\ket}{\bra n1\ket}\tilde\lam_s+\tilde\lam_1.
\end{equation}
The three point amplitude is non-zero when $h_1=-h_I$ and both choices give the same contribution, combining the terms the BCFW recursion gives
\begin{equation}
 M_{n+1}(s,1,\dots,n)=\frac{\bra n1\ket}{\bra ns\ket\bra s1\ket}M_n(-I,2,\dots,n(z^*))+\dots.
\end{equation}
Now reescaling $\lam_s\rightarrow\eps\lam_s$ the above becomes
\begin{align}
&M_{n+1}(\{\eps\lam_s,\tilde\lam\},\{\lam_1,\tilde\lam_1\},\dots,\{\lam_n,\tilde\lam_n\})\\
&=\frac{1}{\eps^2}\frac{\bra n1\ket}{\bra ns\ket\bra s1\ket}M_n(\{\lam_1,\tilde\lam_1+\eps\frac{\bra ns\ket}{\bra n1\ket}\tilde\lam_s\},\lam_2,\tilde\lam_2\},\dots,\{\lam_n,\tilde\lam_n+\eps\frac{\bra s1\ket}{\bra n1\ket}\tilde\lam_s\})\nonumber
\end{align}
$M_n$ is finite when $\eps\rightarrow0$ and corresponds to the stripped $n$ point amplitude. Restore the momentum conserving delta functions and expand the amplitude to first order around $\eps=0$
\begin{align}
&\M_n(\{\lam_1,\tilde\lam_1+\eps\frac{\bra ns\ket}{\bra n1\ket}\tilde\lam_s\},\lam_2,\tilde\lam_2\},\dots,\{\lam_n,\tilde\lam_n+\eps\frac{\bra s1\ket}{\bra n1\ket}\tilde\lam_s\})\\
&=\ld1+\eps\frac{\bra ns\ket}{\bra n1\ket}\tilde\lam_s\cdot\frac{\pd}{\pd\tilde\lam_1}+\eps\frac{\bra 1s\ket}{\bra 1n\ket}\tilde\lam_s\cdot\frac{\pd}{\pd\tilde\lam_n}\rd \M_n(\eps=0),\nonumber
\end{align}
multiplying by the soft factor we arrive at the expression
\begin{align}
&\M_{n+1}(\{\eps\lam_s,\tilde{\lam}_s\},1,\dots,n)=\ld\frac{1}{\eps^2}S^{(0)}+\frac{1}{\eps}S^{(1)}\rd \M_n(1,\dots,n)+\O(\eps^0)\\
&S^{(0)}=\frac{\bra n1\ket}{\bra ns\ket\bra s1\ket},\;\;S^{(1)}=\frac{1}{\bra s1\ket}\tilde\lam_s\cdot\frac{\pd}{\pd\tilde\lam_1}+\frac{1}{\bra ns\ket}\tilde\lam_s\cdot\frac{\pd}{\pd\tilde\lam_n}
\end{align}
as claimed. Notice that there is no sub-sub-leading divergence in this limit as is the case in gravity, here this term would be finite and would mix with the rest of the BCFW recursion terms. Another important difference is that
while these factor are universal for tree amplitudes as in the case for gravity, at loop-level the soft factor receives corrections \cite{Dixon:2008gr,Dixon:2009ur} and so it is expected that also the sub-leading divergence also will receive quantum corrections.

Nevertheless this tree-level data is still interesting, the sub-leading divergences are proportional to the angular momentum operator just like in gravity and might be derived from asymptotic methods as was done for the usual
soft limit in \cite{Strominger:2013lka}. On the other hand this sub-leading divergence is gauge-invariant by itself, not requiring conservation of linear or angular momentum much like the sub-sub-leading divergence of the soft
limit in gravity, it would be interesting to investigate the extent to which these terms are in any sense universal or connected to properties of the asymptotic boundary of flat space. 

\section{Acknowledgements}

I'm grateful to  D. Skinner and T. Adamo for discussions and for D. Skinner for suggesting this problem. This work is supported by the Cambridge Commonwealth, European and International Trust.

\bibliographystyle{utphys}
\bibliography{References}
\end{document}